\documentclass[% 
reprint,
 amsmath,amssymb,
 aps,
pra,
]{revtex4-1}

\newcommand{\entmin}[0]{\ensuremath{H_\text{min}}} %min-entropy
\newcommand{\trace}{\ensuremath{\text{Tr}}}

\newcommand{\bitstring}[1]{\ensuremath{\mathbf{#1}}}

\newcommand{\checkEC}[0]{\ensuremath{\text{check}_\text{EC}}}
\newcommand{\lossPA}[0]{\ensuremath{\text{loss}_\text{PA}}}
\newcommand{\eps}[2]{\ensuremath{\varepsilon^{#1}_\text{#2}}}

\usepackage[english]{babel}
\usepackage[utf8x]{inputenc}
\usepackage[T1]{fontenc}

\usepackage{amsmath}
\usepackage{graphicx}
\usepackage[colorinlistoftodos]{todonotes}
\usepackage[colorlinks=true, allcolors=blue]{hyperref}
\usepackage{graphicx,amssymb,amsmath,amsthm,mathtools,mathrsfs,hyperref}
\usepackage{graphicx}% Include figure files
\usepackage{dcolumn}% Align table columns on decimal point
\usepackage{bm}% bold math
\begin{document}

\preprint{APS/123-QED}

\title{Finite-key security for quantum key distribution systems utilizing weak coherent states}% Force line breaks with \\

\author{Anton Kozubov}
  \email{avkozubov@corp.ifmo.ru}%Lines break automatically or can be forced with \\
\author{Andrei Gaidash}%
\author{George Miroshnichenko}
\affiliation{%
 Department of Photonics and Optical Information Technology, ITMO University,199034 Kadetskaya Line 3b, Saint Petersburg, Russia
}%
\affiliation{Quanttelecom Ltd., Saint Petersburg, 199034 Birzhevaya Line 14, Russia.}

\date{\today}% It is always \today, today,

\begin{abstract}
In this paper we present finite-key security analysis for quantum key distribution protocol based on weak coherent (in particular phase-coded) states using a fully quantum asymptotic equipartition property  technique. This work is the extension of the proof for non-orthogonal states on the coherent states. Below we consider two types of attacks each of them maximizes either Alice-Eve or Eve-Bob mutual information. The cornerstone of this paper is that we do assume the possibility of crucial intercept-resend attack based on errorless unambiguous state discrimination measurement. We demonstrate that Holevo bound always gives the highest mutual information between Alice and Eve regardless particular kind of isometry. As the main result we present the dependence of the extracted secret key length. As the example we implement the proposed analysis to the subcarrier wave quantum key distribution protocol.

\end{abstract}

\maketitle

%\tableofcontents

\section{Introduction}

Growing interest to quantum key distribution (QKD) systems \cite{scarani2009security,diamanti2016practical} in the last decades has led to emergence of a large number of experimental works dedicated to the development of reliable QKD setups suitable for everyday operation in existing telecommunication networks. %Among them stand subcarrier wave (SCW) QKD systems \cite{merolla1999single,merolla2002integrated,mora2012simultaneous,guerreau2005quantum,gleuim2017sideband,gleim2016secure,Miroshnichenko18, gaidash2019methods}, whose the most valuable feature is exceptionally efficient use of quantum channel bandwidth and capability of signal multiplexing by adding independent sets of quantum subcarriers around the same carrier wave \cite{mora2012simultaneous}. It makes SCW QKD systems perfect candidates as a backbone of multiuser quantum networks.

Nevertheless security proofs for different QKD systems still require special consideration. Protocols based on phase-coded coherent states, e.g. in~\cite{Miroshnichenko18}, can still be the good practical solution due to their utility. Despite the unconditional security proofs for non-orthogonal states have already been presented \cite{tamaki2003unconditionally,christandl2004generic,renner2005information}, they still do not cover the case when weak coherent states are used. The differences between approaches are crucial and, from our point view, the paper directs to more practical cases. Hence the protocols based on coherent states are now might be secure by introducing the first analysis of the finite-key security against general attacks.

We use common approach based on assumption that all rounds of protocols are independent and identically-distributed (i.i.d.) \cite{tomamichel2009fully, scaranirenner2008security,scaranirenner2008quantum, lucamarini2013efficient}. This assumption naturally takes place since each of the $N$ signals is prepared, sent, and received independently of the other signals (this is equivalent that the protocol is invariant under permutations of the states held by Bob after the distribution phase\cite{renner2009finetti}).
%\textcolor{magenta}{ which is relatively fair in terms of SCW QKD state representation and its natural symmetry. It is well-known that coherent attacks in i.i.d. case can be bounded with collective attacks. So we consider in this paper coherent attacks as general collective errorless attacks in terms of arbitrary unitary operations on purified states in enlarged Hilbert space (described in terms of isometry) provided by Eve.}

The paper is organized as follows. Section~\ref{extension of the security for B92-like protocols} gives the detailed description of  the representation of the state space and the strategy of possible attacks with maximization over mutual information between Alice and Eve (independent from error and detection rates) on weak coherent (non-orthogonal) states protocols in approximation of infinite number of rounds (in terms of complitely-positive trace-preserving (CPTP) maps), as well as potential vulnerability (consequent from the utilization of weak coherent state) and its practical solution.  Section~\ref{Security notation} gives the detailed description of the finite-key analysis for considered type of protocols.  In Section~\ref{Results} we assert the main results of our paper. In Appendix~\ref{security} we provide the example of the proof implemented to real subcarrier wave (SCW) QKD system.

\section{extension of the security for B92-like protocols}\label{extension of the security for B92-like protocols}
In this section we would like to explain which extensions compared to existing security proofs \cite{tamaki2003unconditionally, christandl2004generic, renner2005information} we present in this paper. Despite the techniques presented in those papers are strong, they are rather impractical due to some assumptions. One of them is requirement of the true single-photon sources while the majority of QKD systems nowadays are based on coherent states (or mixed states) and those security proofs can not be implemented directly to them. Moreover in the papers authors assume that the measurement on the Bob's (and/or Eve's) side is realized in Breidbart basis \cite{bennett1992experimental}. This means that the distinguishing measurement of non-orthogonal states is performed with only (optimized) error and without inconclusive result. Nevertheless the construction of measurement device incarnating this measurement in real live is impossible due to imperfections of the quantum optical devices; they always provide inconclusive results. 

Worth noticing that since we are getting rid of these assumptions the errorless unambiguous state discrimination attack can be provided by Eve.  Thus the extension of the security proof on the case of real systems implementing weak coherent (non-orthogonal) states looks natural.

\subsection{Description of the state space}
First of all let us introduce the space of the states prepared by Alice. In general case the space of the states spaned over the vectors $|\alpha_i\rangle$. Index $i$ denotes different states with arbitrary way of information coding, i.e. phase-coded states. They can be described in terms of representation basis of abelian cyclic point symmetry groups $C_{M}$ respectively (equally distributed in phase plane and with equal \textit{a priori} sending probabilities). Representation of this group in basis $|\alpha_j\rangle$, where $j=1...M$ are denoted by operators $\mathcal{G}_{ph}$, which applied to basis elements as follows 
\begin{gather}\label{U}
\mathcal{G}_{ph}|\alpha_j\rangle=\exp\left(\frac{i 2\pi}M a^\dagger a\right)|\alpha_j\rangle=\Big|\alpha_j \exp\left(\frac{i 2\pi}M\right) \Big\rangle,
\end{gather}
where $a$ is the photon annihilation operator and $M$ is the number of implemented states.

The Hilbert space of the prepared sequences of the states $\mathcal{H}^{\otimes N}$ space, where $N$ is the number of prepaired states and $\mathcal{H}$ is the Hilbert space of each M-dimensional signal. The Hilbert space of the prepared sequences has a natural orthogonal decomposition according to the irreducible representations of the permutation group of N points, acting as the permutations of the tensor factors.
On the space $\mathcal{H}^{\otimes N}$ the representation of the symmetric group of permutations of the order of signals and the tensor representation of the abelian group $C_M$ can be realized. The dimension of this representation is $M^{N}$.
 These two representations are ‘‘commutants’’ of each other according to~\cite{simon1996representations}. Thus the representation of the states allows to symmetrize the space of sequences. 

Since we assume that the considered states are i.i.d., the analysis of coherent attacks can be reduced to the analysis against collective attacks (however the proof of the fact that considered states are i.i.d. is out of scope in this paper). 

Considering collective attack Eve can measure the intercepted states after reconciliation. Such opportunity leads to some crucial advantage. Due to the symmetric structure of the states and their invariance to the unitary rotation, described as an evolution operator, Eve can simply shifts the phase or polarisation of the states by using an appropriate unitary rotation of state. The unitary rotation does not change the accessible information, thus this is equivalent that Eve distinguishes between only two states as in the B92-like protocol case and it does not depend from the number of used states (for the purposes of estimation mutual information between Alice and Eve).

\subsection{Capacity estimation for quantum channel}
\begin{figure}[tp]
\begin{center}
\includegraphics[width=0.9\linewidth]{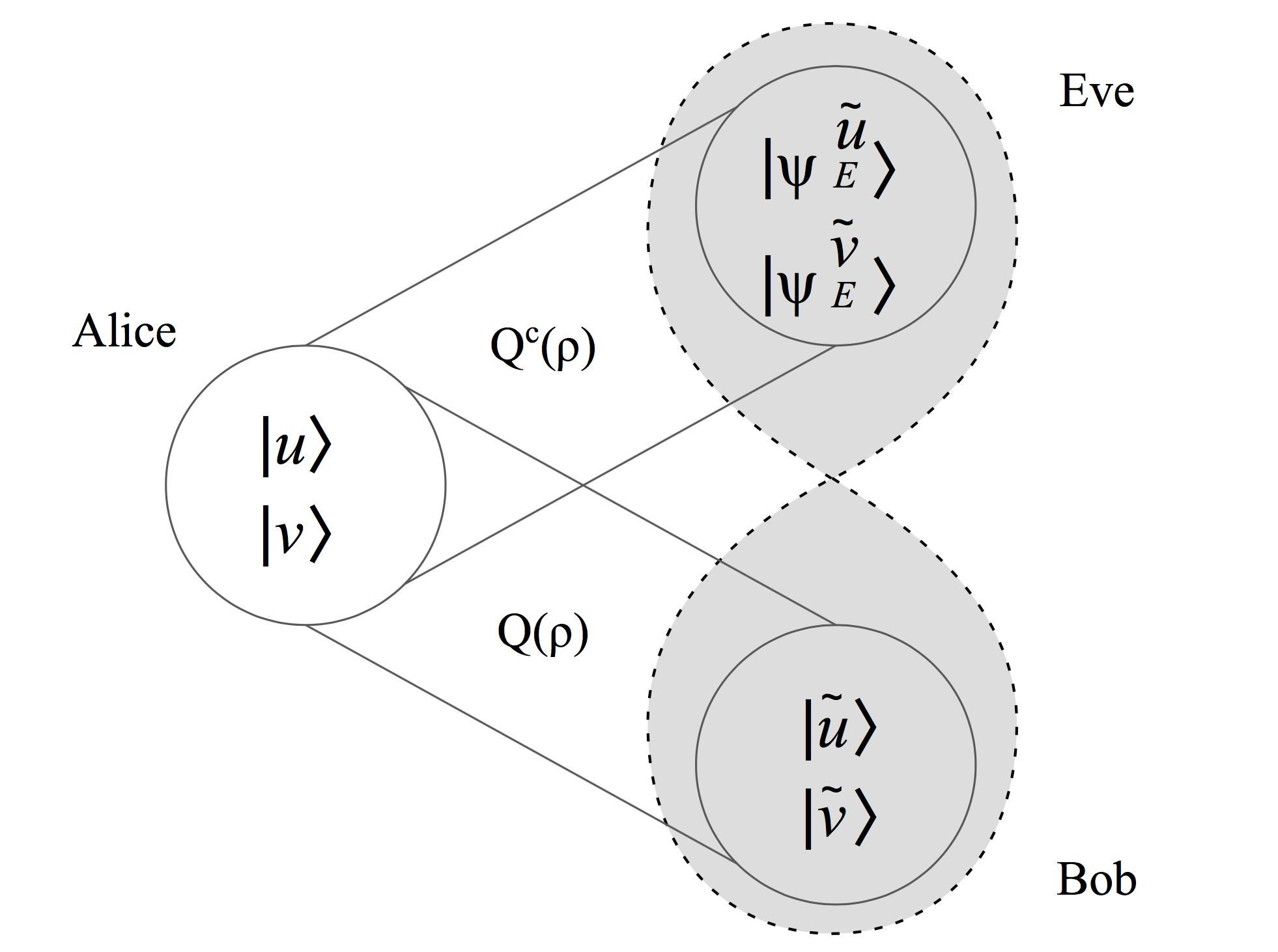} 
\end{center}
\caption{Visual interpretation of considered isometry where initial states $|u\rangle$ and $|v\rangle$ are mapped by (meaning transmission in) quantum channel $\mathcal{Q}(\rho)$ and complimentary quantum channel $\mathcal{Q}^c(\rho)$ to states $|\tilde{u}\rangle$ and $|\tilde{v}\rangle$ available to Bob and states of Eve's ancillas $|\psi_E^{\tilde{u}}\rangle$ and $|\psi_E^{\tilde{v}}\rangle$ respectively. Grey area means possible entanglement of the states or at least their previous interaction. Here $|\psi_E^{\tilde{u}}\rangle$ and $|\psi_E^{\tilde{v}}\rangle$ entangled with $|\tilde{u}\rangle$ and $|\tilde{v}\rangle$ respectively.}
\label{fig:isometry}
\end{figure}

Let us consider the quantum channel as CPTP linear map. There are three different notations known as Stinespring theorem \cite{stinespring1955positive}, Kraus operator, and Choi-Jamiołkowski representation \cite{sutter2017approximate,winter2016potential}.
We will consider Stinespring notation (in terms of unitary isometries, see Fig.~\ref{fig:isometry}) to estimate amount of information leaked to Eve considering complementary channel following \cite{winter2016potential}. After we define bound on Eve's information independent on type of isometry.

Let $\mathcal{A}, \mathcal{B}, \mathcal{E}$ be Hilbert spaces of Alice, Bob and Eve respectively. Stinespring theorem states that for every quantum channel $\mathcal{Q}$ there always exists an isometry $U:\mathcal{A}\hookrightarrow \mathcal{B}\otimes \mathcal{E}$, such that $U^\dagger U=\mathbb{I}$, that maps  from $\mathcal{A}$ to the joint system $\mathcal{B}\otimes \mathcal{E}$. Taking partial trace on the system $\mathcal{E}$  such that $\mathcal{Q}(\rho)=\trace_\mathcal{E} U\rho U^\dagger$ for all $\rho$ belongs to space of trace class
operators acting on Hilbert spaces $\mathcal{A}$. Tracing out system $\mathcal{B}$ instead of $\mathcal{E}$ defines a complementary channel $\mathcal{Q}^c(\rho)=\trace_\mathcal{B} U\rho U^\dagger$. Capacity (more precisely Holevo capacity) of the complementary channel characterizes mutual information; considering it one may derive conditional von Neumann entropy between Alice and Eve (as it will be used further in Eq.~\ref{aep}).
In general case the capacity of quantum complementary channel in terms of QKD can be described as classical capacity $\mathit{C}(\mathcal{Q}^c)$ \cite{holevo1998capacity, schumacher1997sending} and can be denoted as the rate of reliably sent classical information through any quantum channel. It can be expressed using the Holevo capacity $\chi (\mathcal{Q}^c)$ for given state (and given type of isometry) and has the next form assuming i.i.d. case  
\begin{equation}
\mathit{C}(\mathcal{Q}^c)=\lim_{N\rightarrow\infty}\frac{1}{N}\chi ((\mathcal{Q}^{c})^{\otimes n})=\chi (\mathcal{Q}^{c}),
\end{equation}
where $N$ is the number of sent qubits and
\begin{equation}
\chi (\mathcal{Q}^c)=S\left(\sum_x p_x \mathcal{Q}^c(\rho_x) \right)-\sum_x p_x S(\mathcal{Q}^c(\rho_x)),
\end{equation}
where $S(\rho)=-\trace\{\rho \log\rho\}$ is the von Neumann entropy and $\rho$ is the unconditioned channel density operator.

Let us consider the result of isometry in order to estimate Holevo capacity in complimentary channel. We denote states prepared by Alice as
\begin{gather}
    |u\rangle=\mathcal{G}^0_{ph}|\alpha_0\rangle=\mathbb{I}_2|\alpha_0\rangle=|\alpha_0\rangle,\\
    |v\rangle=\mathcal{G}_{ph}|\alpha_0\rangle=\mathcal{G}_{ph}|u\rangle=|-\alpha_0\rangle,
\end{gather}
where $|\alpha_0\rangle$ is the initial state.
Eve performs unitary operation (described by isometry) between states in the channel and Eve's ancillas to make them (in general case) entangled in some way \cite{ekert1994eavesdropping}

\begin{eqnarray}
\begin{cases}
|u\rangle\longrightarrow a|\tilde{u}\rangle\otimes|\psi_\mathcal{E}^{\tilde{u}}\rangle+ b|\tilde{v}\rangle\otimes|\psi_\mathcal{E}^{\tilde{v}}\rangle\\
|v\rangle\longrightarrow b|\tilde{u}\rangle\otimes|\psi_\mathcal{E}^{\tilde{u}}\rangle+a|\tilde{v}\rangle\otimes|\psi_\mathcal{E}^{\tilde{v}}\rangle,
\end{cases}
\end{eqnarray}

where $a$ and $b$ are arbitrary coefficients satisfying norm and unitarity conditions, $\{|u\rangle,|v\rangle\}$ is the space of prepared by Alice states, $\{|\tilde{u}\rangle,|\tilde{v}\rangle\}$ is the space of states in quantum channel after Eve's actions, $\{|\psi_\mathcal{E}^{\tilde{u}}\rangle,|\psi_\mathcal{E}^{\tilde{v}}\rangle\}$ is the space of Eve's ancillas in complimentary channel (one may assume without loss of generality that Eve had the initial state $|\psi_\mathcal{E}\rangle$ of her ancilla considering unitary evolution of purificated state).

It can be shown that Holevo capacity of complementary channel is maximized when $a=1$ and $b=0$ (or vise versa) impying untangled (but interacted) states

\begin{eqnarray}
\begin{cases}
|u\rangle\longrightarrow|\tilde{u}\rangle\otimes|\psi_\mathcal{E}^{\tilde{u}}\rangle\\
|v\rangle\longrightarrow|\tilde{v}\rangle\otimes|\psi_\mathcal{E}^{\tilde{v}}\rangle.
\end{cases}
\end{eqnarray}

It is known that isometry preserve the overlappings of the states
\begin{eqnarray}
\langle u|v\rangle = \langle\tilde{u}|\tilde{v}\rangle \cdot \langle\psi_\mathcal{E}^{\tilde{u}}|\psi_\mathcal{E}^{\tilde{v}}\rangle. 
\end{eqnarray}
In order to gain any information Eve should (unambiguously or with error) distinguish between $|\psi_\mathcal{E}^{\tilde{u}}\rangle$ and $|\psi_\mathcal{E}^{\tilde{v}}\rangle$ hence these states should be separated, i.e. $\langle\psi_\mathcal{E}^{\tilde{u}}|\psi_\mathcal{E}^{\tilde{v}}\rangle<1$, thus $\langle\tilde{u}|\tilde{v}\rangle\le1$.
Thus one may deduce the following statement implying Eve has any information
\begin{equation}\label{ineq}
\langle u|v\rangle \le \langle\psi_\mathcal{E}^{\tilde{u}}|\psi_\mathcal{E}^{\tilde{v}}\rangle.
\end{equation}

Holevo capacity is decreasing function of overlapping. Hence one may substitute left-hand side for right-hand side overlapping in Eq.~\ref{ineq} as new argument of Holevo capacity. This gives us higher information bound (the Holevo bound) and also eleminates the neccessity to consider particular kinds of isometries

\begin{eqnarray}
\chi (\rho) \ge \chi (\mathcal{Q}^c(\rho)).
\end{eqnarray}

Holevo bound depends on mean photon number (as well as overlapping of the states), hence it can reach the unity for higher intensities. Due to this one should take into consideration only the practical cases when mean photon number is low enough. 

\subsection{Potential vulnerability}
 However consideration of only the Holevo bound (which is independent from error and detection rate) due to implementation of the weak coherent states is insufficient and leads to crucial loophole covered further. As far as weak coherent states are also overlapped, they cannot be distinguished perfectly, although Eve can discriminate them unambiguously only with the some probability (depending on the used states and hence the values of their overlapping). In comparison to the security proof for single-photon QKD schemes the unambiguous state discrimination (USD) attack with inconclusive result yet zero error can be implemented.
 
 In this work we consider only errorless USD attack since it is much more crucial then discrimination with optimized error (Helstrom bound \cite{helstrom1976quantum}). The construction of the USD positive-operator valued measure (POVM) provides for Eve only two possible outcomes for each state: the unambiguous identification of the particular state and the inconclusive result. All inconclusive results can be blocked (exchanged with vacuum states) by Eve and she can increase the intensity of sent pulses to maintain the detection rates. It is possible due to the big number of inconclusive results on the Bob's side due to losses in the channel and low quantum efficiency of detectors. 
 
 The strategy allows Eve to provide zero-error attack keeping up to 100\% of shared bits between Alice and Bob and maintaining both raw key rates and error rates. Surprisingly the attack is more crucial than some collective attacks (for instance collective beam-splitting attack in \cite{Miroshnichenko18}). It can be explained by the fact that the information obtained by collective attacks on the systems with weak coherent states is mutual information between Alice and Eve $I(A;E)$ and it can be bounded with Holevo bound. Otherwise when we analyse the USD attack we have to estimate mutual information between Eve and Bob $I(E;B)$. In case of the attack the latter mutual information is higher than Holevo bound and equals to unity since all quantum bits detected by Bob are send by Eve. Due to the fact there is no possibility to extract any secret key. 

\subsection{Practical solution}\label{solution}
Nevertheless there is solution providing practical countermeasures against the attack. In paper \cite{gaidash2018overcoming} it was shown that if we satisfy the next condition Eve will be revealed
\begin{equation}\label{usdcond}
    1-G>P_{USD},
\end{equation}
where $1-G$ is expected detection probability (by $G$ we assume the probability of "null" count, i.e. see Eq.~\ref{detprob} in the Appendix as an example) and $P_{USD}$ is probability of unambiguous state discrimination. Obviously there are two main strategies. The first is to increase $1-G$, the second is to decrease $P_{USD}$. The second strategy looks more convenient from the practical point of view  (inequality~\ref{usdcond} should be satisfied for all acceptable channel losses where we can extract secret key). To do so one can simply increase the number of states (bases). If following inequality satisfies that there might be USD attack with the probability erf$(\frac{z}{\sqrt{2}})$, where the latter is error function,
\begin{gather}\label{detectionrate}
n<N\cdot (1-G)-z\sigma,\\
\sigma=\sqrt{N\cdot G (1-G)},
\end{gather}
where $n$ is the number of detected quantum bits, $N$ is the number of sent signals (its value should be chosen appropriately \cite{gaidash2018overcoming}) and $z$ is the arbitrary number of standard deviations $\sigma$ within the confidence interval according to the so-called "three--sigma rule". If the inequality satisfies we abort the protocol. However there is arbitrary small probability $1-$erf$(\frac{z}{\sqrt{2}})$ that USD attack would not be detected.

\section{Security notation}\label{Security notation}
After the quantum states are sent and measured, Alice and Bob each have $n$-bit strings $\bitstring{A}$ and $\bitstring{B}$. Correspondingly Eve has side information $\bitstring{E}$. To estimate appropriate bound on secure key rate we consider the notation of Renyi entropies $H_{\alpha}(X)=\frac{1}{1-\alpha}\log\left(\sum_{i=1}^n p_i^\alpha\right)$, since they describe the worst case and not the average one. In the paper we consider that $\alpha\rightarrow\infty$ since we use \textit{min-entropy} $H_\infty(X)=\entmin =-\log \max_i p_i$. Futher we will use quantum asymptotic equipartition property (QAEP) \cite{tomamichel2009fully}. It means that for a large number of rounds, the operationally relevant total uncertainty can be well approximated (and bounded) by the sum over all independent and identically-distributed (i.i.d.) rounds. Thus we bound \emph{$\varepsilon$-smooth min-entropy} \cite{renner2005security,tomamichel2012framework} as follows
\begin{equation}\label{aep}
\entmin^{\varepsilon_S}(\bitstring{A}|\mathbf{E})\geq n\left(H(\bitstring{A}|\mathbf{E})-\frac{\delta(\varepsilon_S)}{\sqrt{n}}\right),
\end{equation}
where 
\begin{equation}
\delta(\varepsilon_S)=4\log(2+\sqrt{2})\sqrt{\log\left( \frac{2}{\varepsilon_S^2} \right)}
\end{equation}
and $H(\bitstring{A}|\mathbf{E})$ is conditional von Neumann entropy and it denotes the entropy of Alice’s bit conditioned on Eve’s side-information in a
single round (and it is constant due to i.i.d. assumption), in terms of coherent-state-based protocols it can be bounded as $H(\bitstring{A}|\mathbf{E})\geq 1-\chi(\rho)$, where $\rho$ is unconditioned channel density operator, which can be expressed as follows
\begin{eqnarray}
\rho=\sum_i p_i \rho_i
\end{eqnarray}
where $\rho_i$ is the density matrix of each state, and $p_i$ is the \text{a priori} sending probability of this state. 
\subsection{Parameter estimation}
Nevertheless we can apply proposed analysis to the protocol only after parameter estimation, since we still have to reveal any intercept-resend attacks. Exchanging the conditional von Neumann entropy with Holevo bound which is constant and independent on error or detection rate is a crucial point there. Performing intercept-resend attack with error on small fraction of qubits does not give Eve more information than Holevo bound otherwise it will produce too high QBER leading to negative secure key rates. In other words there is no common parameter estimation on QBER. Its not the only parameter that we should worry about due to introduced errorless attack. Hence in case of errorless intercept-resend attack described above one should monitor the detection rates according to the inequality~\ref{detectionrate}. If the inequality is satisfied one should abort the protocol, on the contrary the following security analysis can be applied.

\subsection{Error correction}
On this step both parties should check and correct the errors in their bit strings. It can be done using any error correction code. In the paper we assume that Alice and Bob uses low-density parity-check (LDPC) codes at the error correction step. Bob publicly sends a random subset of $k$ bits to Alice, and she estimates quantum bit error rate (QBER) $Q_{est}$ in that subset. It should be noted that LDPC codes succeed if value of real QBER $Q_{real}$ (see Appendix~\ref{Detection scheme} for more details) less than reference value parametrized the code. Thus Alice needs additional QBER $\Delta Q$. For instance it can be estimated in order to maximize probability of successful error correction in one round and keeping maximal secret key rate. It can be found considering bounds for sampling without replacement \cite{serfling1974probability} as follows
\begin{equation}\label{deltaq}
Pr(Q_{est}+\Delta Q<Q_{real})\le \exp{\left( \frac{-2k\Delta Q^2}{1-(k-1)/n} \right)},
\end{equation}
where \textit{k} is the number of bits in the sample. Then Alice computes the syndrome of LDPC code that corrects up to $n(Q_{est}+\Delta Q)$ error bits. We denote the length of the syndrome as $code_{EC}$ which is some function of $n(Q_{est}+\Delta Q)$. For (only) analytical and/or estimation purposes we may express $code_{EC}$ as asymptotic approximation
\begin{equation}\label{codeecap}
code_{EC} \approx n f_{EC} h(Q_{est}+\Delta Q),
\end{equation}
where $f_{EC}$ is error correction efficiency. Using the syndrome, Bob corrects his bits to some new bitstring $\bitstring{B'}$. Then Bob applies a 2-universal hash function with output length $check_{EC}$ to $\bitstring{B'}$ and sends hash to Alice checking whether their strings match. If hashes are different Alice enlarges $\Delta Q$ or abort the protocol. Otherwise Alice obtained the bitstring $\bitstring{A'}$. The remaining smooth-entropy is conservatively lower-bound by simply subtracting off the total number of bits publicly revealed
\begin{eqnarray}
\begin{split}
\entmin^{\varepsilon_S}(\bitstring{A'}|\mathbf{E})\geq n\left(H(\bitstring{A}|\mathbf{E})-\frac{\delta(\varepsilon_S)}{\sqrt{n}}\right)- \\
-k-code_{EC}-check_{EC},
\end{split}
\end{eqnarray}
We would like to stress out that the value $code_{EC}$ put into this bound must be the number of bits $actually$ sent in particular round, approximation in Eq.~\ref{codeecap} may be used only for analytical and/or estimation purposes.
\subsection{Privacy amplification and key extraction}\label{Security parameters}
At this point Alice and Bob performs privacy amplification. They hash their bitstrings to a key of length $l$ (see \cite{tomamichel2011leftover}), where
\begin{equation}
\begin{split}
l = n\left(H(\bitstring{A}|\mathbf{E})-\frac{\delta(\varepsilon_S)}{\sqrt{n}}\right)-k- \\
-code_{EC}-check_{EC}-loss_{PA},
\end{split}
\end{equation}
where $loss_{PA}$ is a parameter chosen in order to achieve a curtain level of security (see further in following subsection).

We use the definitions given in Section 2.5 of \cite{arnon2016simple} and roughly follow the analysis in Section 5.1.

When considering the error correction step, we have to estimate ``correctness error'' $\eps{}{EC}$. Alice and Bob compare hashes of length $\checkEC$ ensures that the probability that $\bitstring{A}' \neq \bitstring{B}'$ \emph{and} Alice and Bob fail to abort (i.e. their hashes match) is
\begin{align}
\eps{}{EC} = 2^{-\checkEC}.
\label{eq_checkEC}
\end{align}
This follows from the properties of 2-universal hashing.

For the privacy amplification, the output of the extractor has length $l \leq \entmin^{\varepsilon_s}(\bitstring{A}'|\mathbf{E}) - \lossPA$. Therefore, using the bound in \cite{tomamichel2011leftover} then tells us that the trace distance $d$ between the protocol's output and an ideal output (where the key is uniform and independent from Eve, even after Eve knows the matrix used for the hashing) is bounded above by  
\begin{equation}
\begin{split}
d &= \frac{1}{2}\Vert \rho_{KFE}-\omega_{K}\otimes\sigma_{FE}\Vert_{1} \\
&\leq \varepsilon_s + \frac{1}{2}\sqrt{2^{l - \entmin^{\varepsilon_s}(\bitstring{A}'|\mathbf{E})}} \\
&\leq \varepsilon_s + \frac{1}{2}\sqrt{2^{-\lossPA}} \label{eq_epssec} \\
&\leq \varepsilon_s + \eps{}{PA} =\varepsilon_{sec},
\end{split}
\end{equation}
where in the last step we introduce the quantity $\varepsilon_{sec}$ as an upper bound on the trace distance $d$.

Overall, in terms of the definitions given in Section 2.5 of \cite{arnon2016simple}, this means that the protocol is $\varepsilon_{corr}$-correct with $\varepsilon_{corr} = \eps{}{EC}$ (see Definition 1) and $\varepsilon_{sec}$-secure with $\varepsilon_{sec} = \varepsilon_s + \eps{}{PA}$ (see Definition 2). Phrasing it in terms of Definition 3, the protocol is hence $\varepsilon_\text{QKD}$-secure-and-correct, with $\varepsilon_\text{QKD} = \eps{}{EC} + \varepsilon_s + \eps{}{PA}$ providing secure bit string with length
\begin{equation}\label{lqkd}
\begin{split}
l = n(1-\chi(\rho))-4\sqrt{n}\log(2+\sqrt{2})\sqrt{\log\left( \frac{2}{\varepsilon_S^2} \right)}- \\
-k-code_{EC}-\log\frac{1}{\eps{}{EC}}-\log\frac{1}{\eps{}{PA}}+2.
\end{split}
\end{equation}

\section{Results}\label{Results}
In this paper we consider security of coherent-state-based QKD protocols. We discuss the differences between existing proofs and the one in present paper and the consequences of this extension. The cornerstone of this work is that we do assume the possibility of crucial intercept-resend attack based on errorless USD measurement. It is justified by the fact that weak coherent states are used and the measurement at the Bob's side always produces inconclusive result. 

Introduction of Holevo bound and alternation of parameter estimation cover security loopholes. The inequality for detection rate estimation introduced in Section~\ref{solution}. We demonstrate in details why we use Holevo bound as the highest amount of mutual information between Alice an Eve regardless particular kind of attack described in terms of isometry. Finite-key security analysis is performed by implementation of fully quantum asymptotic equipartition property. As the main result we estimate extracted secret key length (Eq.~\ref{lqkd}). In the appendix we present the implementation of proposed proof for subcarrier wave (SCW) QKD system. The average secure key rate $R$ dependence on channel losses in SCW QKD system for different number of detected quantum bits is presented in Appendix~\ref{Appendix B: Performance estimation}.

\section*{Acknowledgements}
This work was financially supported by Government of Russian Federation (Grant 08-08).
All authors contributed equally to the work.
Also we are very grateful to Ernest Tan and Renato Renner for fruitful discussions on finite-key analysis and to Vera Shurygina for inestimable help with figure preparation.
\appendix

\section{Security analysis for SCW QKD protocol}\label{security}
In the appendix we would like introduce the example of such security proof for SCW QKD protocol. One of the most valuable feature of SCW QKD systems \cite{merolla1999single,merolla2002integrated,mora2012simultaneous,guerreau2005quantum,gleuim2017sideband,gleim2016secure,Miroshnichenko18, gaidash2019methods} is exceptionally efficient use of quantum channel bandwidth and capability of signal multiplexing by adding independent sets of quantum subcarriers around the same carrier wave \cite{mora2012simultaneous}. It makes SCW QKD systems perfect candidates as a backbone of multiuser quantum networks. The following section give the detailed description of the quantum states, Holevo bound for used states, detection scheme and performance estimation. 

\subsection{Quantum state preparation}\label{Quantum state preparation}
Several kind of protocols can be implemented in our system, i.e. with two ($\varphi_A\in\{0,\pi\}$), four ($\varphi_A\in\{0,\pi, \pi/2, 3\pi/2\}$) and, in general, any even number of signal weak coherent phase-coded states which can be described in terms of representation basis of abelian cyclic point symmetry groups $C_M$ respectively (equally distributed in phase plane and with equal \textit{a priori} sending probabilities). We propose here the protocol based on sixteen states (eight bases). The initial state (before modulation) can be described as $|\sqrt{\mu_0}\rangle_0\otimes|\mathrm{vac}\rangle_{SB}$, where $|\mathrm{vac}\rangle_{SB}$ is the vacuum state of the sidebands and $|\sqrt{\mu_0}\rangle_0$ is the coherent state of the carrier wave with the amplitude determined by the average number of photons $\mu_0$ in a transmission window provided with coherent monochromatic light beam with optical frequency $\omega$.  The carrier wave phase is accepted as reference and all other phases are calculated with respect to it. Electro-optical phase modulator (with the frequency of the microwave field $\Omega$ and its phase $\varphi_A$) rearranges the energy between the interacting modes (the field at the modulator output acquires sidebands at frequencies $\omega_k=\omega+k\Omega$, we limit our consideration to $2S$ sidebands and let the integer $k$ run between the limits $-S\le k\le S$), so that the state of the field at the modulator output (in quantum channel) is a multimode coherent state
\begin{equation}\label{phi}
|\psi(\varphi_A)\rangle = \bigotimes_{k=-S}^S|{\alpha_k(\varphi_A)}\rangle_k,
\end{equation}
where amplitudes of coherent states presented as
\begin{equation}\label{alpha}
\alpha_k(\varphi_A)=\sqrt{\mu_0}d^S_{0k}(\beta)e^{-i(\theta_1+\varphi_A)k},
\end{equation}
where $\theta_1$ is a constant phase and $d^S_{nk}(\beta)$ is the Wigner d-function from the quantum theory of angular momentum \cite{varshalovich1988quantum}, $\beta$ is determined by the modulation index $m$, disregarding the modulator medium dispersion the dependence can be written as
\begin{equation}\label{beta}
\cos{({\beta})}=1-\frac{1}{2}{\left(\frac{m}{S+0.5}\right)^2}.
\end{equation}
where $2S+1$ is number of interacting modes and it is considerably large.
The more detailed description of the states and their properties can be found in \cite{Miroshnichenko17,Miroshnichenko18}. 

Under certain conditions, the model of such modulation process is exactly solvable and can be analysed using the technique of the Jordan mappings for the $SU(2)$  Lie algebra \cite{Miroshnichenko17}.

\subsection{Holevo bound}
In case of subcarrier wave quantum key distribution Holevo bound can be found considering unconditioned channel density operator (this is possible due to the fact that Eve can rotate all states stored in her quantum memory after reconciliation and before her measurment)

\begin{equation}\label{rho}
\begin{split}
  \rho=\frac12|u\rangle\langle u|+\frac12|v\rangle\langle v| = \\
  =\frac12|\psi(0)\rangle\langle\psi(0)|+\frac12|\psi(\pi)\rangle\langle\psi(\pi)|,
\end{split}
\end{equation}
where $|\psi(0)\rangle$ and $|\psi(\pi)\rangle$ are denoted according to \ref{phi}.
The von Neumann entropy of a density operator is the Shannon entropy of its eigenvalues. The eigenvalues of the operator $\rho$ are
\begin{equation}\label{lambda}
\lambda_{1,2}=\frac{1}{2}\left(1\pm |\langle\psi(0)|\psi(\pi)\rangle|\right),
\end{equation}
The overlapping of our states can be described as (for more details see \cite{Miroshnichenko18})
\begin{equation}
\langle\psi(0)|\psi(\pi)\rangle=\exp\left[-\mu_0\left(1-d^S_{00}(2\beta)\right)\right].
\end{equation}
Therefore we obtain the Holevo bound using binary Shannon entropy function $h(x)=-x\log x-(1-x)\log(1-x)$ of the eigenvalues defined by Eq.~(\ref{lambda}):
\begin{equation}\label{chi2}
\chi(\rho) = h\left(\frac12(1-\exp\left[-\mu_0 \left(1-d^S_{00}(2\beta)\right)\right]\right).
\end{equation}

\subsection{Detection scheme}\label{Detection scheme}

The main idea of detection scheme is to provide constructive or destructive interference of the whole spectrum of sidebands, thereby distinguishing received states. On the Bob's side there is the phase modulator which has the same frequency $\Omega$ as in the  Alice's one, but a different phase $\varphi$. In this modulator, additional field is generated on the same sidebands $\omega+k\Omega$, which interferes with the field already present on these frequencies. The resulting state of the field is also the multimode coherent state 
\begin{equation}\label{psiB}
|\psi_B(\varphi_A,\varphi)\rangle = \bigotimes_{k=-S}^S|{\alpha_k'(\varphi_A,\varphi)}\rangle_k,
\end{equation}
with coherent amplitudes
\begin{equation}\label{alphaprime}
\alpha_k'(\varphi_A,\varphi)=\sqrt{\mu_0\eta(L)}\exp(-i\theta_2k) d^S_{0k}(\beta'),
\end{equation}
where $\eta(L)=10^{-\xi L/10}$ is transmission coefficient of the quantum channel, $\xi$ is the fiber loss per length unit, and the new argument of the d-function is determined by the relation
\begin{equation}\label{betaprime}
\cos{\beta'}=\cos^2{\beta}-\sin^2{\beta}\cos\left(\varphi_A-\varphi+\varphi_0\right),
\end{equation}
where $\theta_2$ and $\varphi_0$ are some phases determined by the construction of the phase modulator \cite{Miroshnichenko17}.
\begin{figure}[tp]
\begin{center}
\includegraphics[width=\linewidth]{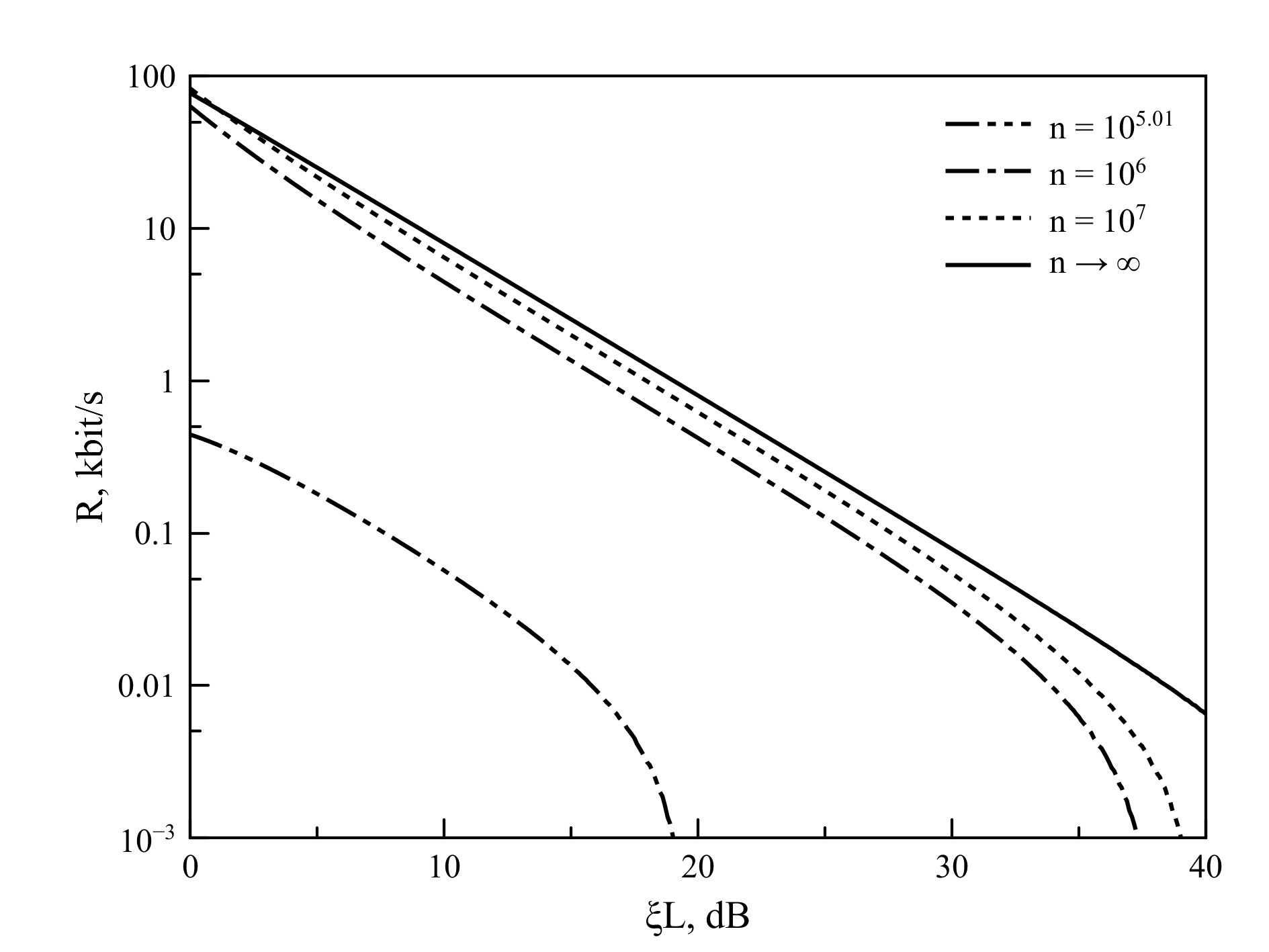}
\end{center}
\caption{Secure key rate $R$ dependence on channel losses in SCW QKD system with different number of detected quantum bits. It should be noted that for $n=10^5$ one cannot generate secure key, and for $n$ slightly more than $10^5$ secure key rate drops drastically compared to close to asymptotic cases.}
\label{fig:rates}
\end{figure}

Equation~(\ref{betaprime}) shows that in order to achieve constructive interference on the sidebands, Bob should use $\varphi_0$ as an offset for his phase and apply in his modulator the microwave phase $\varphi=\varphi_0+\varphi_B$. Then for $\varphi_A-\varphi_B=0$ the argument of d-function doubles: $\beta'=2\beta$ while for $\varphi_A-\varphi_B=\pm\pi$ it vanishes: $\beta'=0$. Since $d^S_{0k}(0)=\delta_{0k}$, a zero argument corresponds to the presence of photons only on the carrier frequency, with all the sidebands being in the vacuum state.

We are interested only in sidebands of our spectrum due to all information about the phase is presented in the sidebands and the carrier wave phase is accepted as reference. To do so we apply spectral filtering in Bob's module, which aims at removing the relatively strong carrier wave. Unfortunately, in a practical SCW QKD system this wave can only be attenuated by a factor $\vartheta\ll1$, resulting in replacement $\bar\alpha_0(\varphi_A,\varphi_B) \to \sqrt{\vartheta}\bar\alpha_0(\varphi_A,\varphi_B)$.

Thus, the average number of photons arriving at Bob's detector in the gate window $T$ in time domain is given by the average total number of photons at all spectral components
\begin{eqnarray}\label{nph}\nonumber
n_{ph}(\varphi_A, \varphi_B)&=& \vartheta|\bar\alpha_0(\varphi_A,\varphi_B)|^2+\sum_{k\ne 0} |\bar\alpha_k(\varphi_A,\varphi_B)|^2 \\
&=&\mu_0\eta(L)\eta_B\left(1-(1-\vartheta)| d^S_{00}(\beta')|^2\right),
\end{eqnarray}
where $\eta_B$ is the losses in Bob's module and according to the property of d-functions \cite{varshalovich1988quantum}
\begin{eqnarray}\label{d-func}
\sum_{k=-S}^S d^S_{nk}(\beta)\left(d^S_{lk}(\beta)\right)^* = \delta_{nl},
\end{eqnarray}
meaning that $d^S_{nk}(\beta)$ is a unitary matrix with respect to its lower indices

Thus probability of a click on the detector operating in Geiger regime in the window $T$ is as follows according to Mandel approximation (considering $n_{ph}\ll 1$)
\begin{equation}\label{pdet}
P_{det}(\varphi_A, \varphi_B)=\left( \eta_D\frac{n_{ph}(\varphi_A, \varphi_B)}{T}+\gamma_{dark} \right)\Delta t
\end{equation}
where $\eta_D$ is the detector quantum efficiency, $\gamma_{dark}$ is the dark count rates, $\Delta t = T$ for continuous operation of the detector, and $\Delta t < T$ is gating time of the detector if there any. Using Eq.~\ref{pdet} one may estimate detection probabilities ($1-G$), error probabilities ($E$), and QBER ($Q$) using  following notation
\begin{gather}
E = P_{det}(0, \pi+\Delta\varphi), \\
1-G = P_{det}(0, \Delta\varphi) + P_{det}(0, \pi+\Delta\varphi), \label{detprob} \\
Q = \frac{E}{1-G}, \label{qber}
\end{gather}
where $\Delta\varphi$ is slight phase instability caused, for instance, by jitter or imperfect synchronization.

\subsection{Performance estimation}\label{Appendix B: Performance estimation}
Let us introduce performance estimations as an example. 
We use $\mu_0=4$ and $m=0.319$ in order to provide optimal mean photon number at sidebands which was found to be $\mu = 0.2$ in \cite{Miroshnichenko18}. Duration of the pulses $T$ is 10 ns (it is equivalent to 100 MHz repetition rates). The fraction of the carrier transmitted through the filter $\vartheta$ is $10^{-3}$, phase instability $\Delta\varphi$ is $5^{∘}$, total losses at the Bob's side $\eta_B$ is 6.4 dB. We refer to the parameters of the detector ID230 from ID Quantique: quantum efficiency $\eta_D$ is $25\%$ and dark counts are at 25 Hz rate. Probability of failing error correction in one round is chosen to be equal to $10^{-6}$ hence the precise value of sample size $k$ is estimated maximizing secret key rate, i.e. minimizing expression $h(Q+\Delta Q(k))+k$ by $k$ (for instance, $k$ is approximately $4\%$ of total received pulses for $n = 10^6$ and approximately $2\%$ of total received pulses for $n = 10^7$). Also security parameter $\varepsilon_\text{QKD}=3\cdot 10^{-10}$ with $\eps{}{EC}=\varepsilon_s=\eps{}{PA}=10^{-10}$.

Finally dividing Eq.~\ref{lqkd} by $n$ to estimate secret key fraction and multiplying it by repetition rate and detection probability in Eq.~\ref{pdet} we obtain dependencies of average secret key rates on losses in the channel for different values of $n$ as follows
\begin{align}
R &= \frac{F(1-G)}{M}\cdot \Big(1-\chi(\rho)- \nonumber \\
&-4\frac{1}{\sqrt{n}}\log(2+\sqrt{2})\sqrt{\log\left( \frac{2}{\varepsilon_S^2} \right)}-\\
&-\frac{1}{n}\big(k+code_{EC}+\log\frac{1}{\varepsilon_{EC}}+\log\frac{1}{\varepsilon_{PA}}-2\big)\Big), \nonumber
\end{align}
where $F$ is the signal sending frequency. Those dependencies for different $n$ are shown in Fig.~\ref{fig:rates}. 
Obtained results are important for constructing secure long-distance QKD links and multiuser quantum networks by exploiting advantages of the SCW QKD: ultra-high QKD bandwidth capacity and compatibility with existing optical communication infrastructure.
\bibliography{bibliography1}

\end{document}